\documentclass[draft]{agujournal2019}
\usepackage{url} 
\usepackage{lineno}
\usepackage[inline]{trackchanges} 
\usepackage{soul}
\draftfalse

\journalname{JGR: Atmospheres}
\begin{document}

\title{Upward Lightning at the Gaisberg Tower: Initiation Mechanism and Flash Type and the Atmospheric Influence}

\authors{Isabell Stucke\affil{1,2}, Deborah Morgenstern\affil{1,2},Gerhard Diendorfer\affil{3}, Georg J. Mayr\affil{2}, Hannes Pichler\affil{3}, Wolfgang Schulz\affil{3}, Thorsten Simon\affil{4}, Achim Zeileis\affil{1}}

\affiliation{1}{Institute of Statistics, University of Innsbruck, Austria, Innsbruck}
\affiliation{2}{Institute of Atmospheric and Cryospheric Sciences, University of Innsbruck, Austria, Innsbruck}
\affiliation{3}{OVE Service GmbH, Dept. ALDIS (Austrian Lightning Detection $\&$ Information System), Austria, Vienna}
\affiliation{4}{Department of Mathematics, University of Innsbruck, Austria, Innsbruck}

\correspondingauthor{Isabell Stucke}{isabell.stucke@uibk.ac.at}

\begin{keypoints}
\item The closer the thundercloud's main electrification region to the tower, the higher the probability that upward lightning is self-initiated.
\item Whether upward lightning is undetectable by lightning location networks can be less reliably explained by larger-scale meteorology.
\item The network-detectable flash type occurs more often accompanied by downward lightning in the vicinity than the undetectable flash type.
\end{keypoints}

----------------------------------------------------------------------- 

\begin{abstract}
Upward lightning is much rarer than downward lightning and requires tall ($100+$~m) structures to initiate. It may be either triggered by other lightning discharges or completely self-initiated. While conventional lightning location systems reliably detect downward lightning, they miss a specific flash type of upward lightning that consists only of a continuous current. Globally, only few specially instrumented towers can detect this flash type. The proliferation of wind turbines in combination with large damage from upward lightning necessitates an improved understanding under which conditions the self-initiated and the undetected subtype of upward lightning occur.
To find larger-scale meteorological conditions favorable for self-initiated and undetectable upward lightning, this study uses a random forest machine learning model. It combines direct measurements at the specially instrumented tower at Gaisberg mountain in Austria with explanatory variables from larger-scale atmospheric reanalysis data (ERA5).
Atmospheric variables reliably explain whether upward lightning is self-initiated by the tower or triggered by other lightning discharges. The most important variable is the height of the $-10~^\circ$C isotherm above the tall structure: the closer it is the higher is the probability of self-initiated upward lightning. Two-meter temperature and the amount of CAPE are also important.
For the occurrence of upward lightning undetectable by lightning location systems, this study finds a strong relationship to the absence of lightning in the vicinity.

\end{abstract}

\section*{Plain Language Summary}
Upward lightning is much rarer than downward lightning. It needs tall structures of $100+$~m to initiate. A large fraction of UL can only be detected by specially equipped towers directly measuring the upward lightning flash. The remaining fraction is undetected by lightning location networks measuring lightning remotely. Further, upward lightning can be initiated by the tall structure itself. Since so many tall wind turbines are built to generate electricity, an improved understanding under which conditions the self-initiated and the undetected subtype of upward lightning occur becomes increasingly important.
Data from one of these specially equipped towers and data from atmospheric conditions are combined in a machine learning model.
Whether lightning strikes in the vicinity of the tower has a strong impact on which flash type occurs. No lightning is more often related to the occurrence of the flash type which is undetectable by the networks.
Meteorological conditions play a minor role. Meteorological conditions, on the other hand, determine, whether UL is self-initiated by the tower. The most important factor is the height at which $-10~^\circ$C are measured above the tower. The closer to the tower, the higher the chance that the tower itself triggers upward lightning.

%
%

\section{Introduction}\label{sec:introduction}

Upward lightning (UL) initiated from the earth surface extending towards the clouds is much rarer than downward lightning initiated within the clouds extending towards the ground. Nevertheless, UL poses a much larger damage potential as it is capable of transferring large amounts of charge up to hundreds of coulombs within a comparably long period of time \cite<e.g.,>[]{diendorfer2015,birkl2017}. Tall structures (on the order of $100$~m) are preferred starting locations for UL \cite<e.g.,>[]{rakov2003}. Wind turbines typically exceed such heights and consequently lightning damages to them have gone up hand in hand with their ever-growing number in the quest for renewable energy sources \cite<e.g.,>[]{rachidi2008,montana2016, birkl2018,pineda2018}. Therefore, the need for a comprehensive risk assessment of damages to wind turbines and other tall structures becomes increasingly relevant.

Proper risk assessment is, however, crucially impeded as more than $50$~\% of UL is not detected by lightning location systems \cite<LLS,>[]{diendorfer2015}. One particular subtype of UL is responsible for the low detection efficiency. It has only a relatively low-amplitude constant initial continuous current  \cite<ICC\textsubscript{only}, e.g.,>[]{birkl2018}. LLS require the fast rising current in the lightning channel to emit sufficiently large electromagnetic field pulses to be detected \cite{diendorfer2009}.  Consequently, using only LLS observations substantially underestimates the risk of lightning damage \cite<e.g.,>[]{rachidi2008,smorgonskiy2013}. Globally, only few specially instrumented towers exist that can measure the ICC\textsubscript{only} UL type.

\begin{figure}
\begin{center}
 \includegraphics[width=8.3cm]{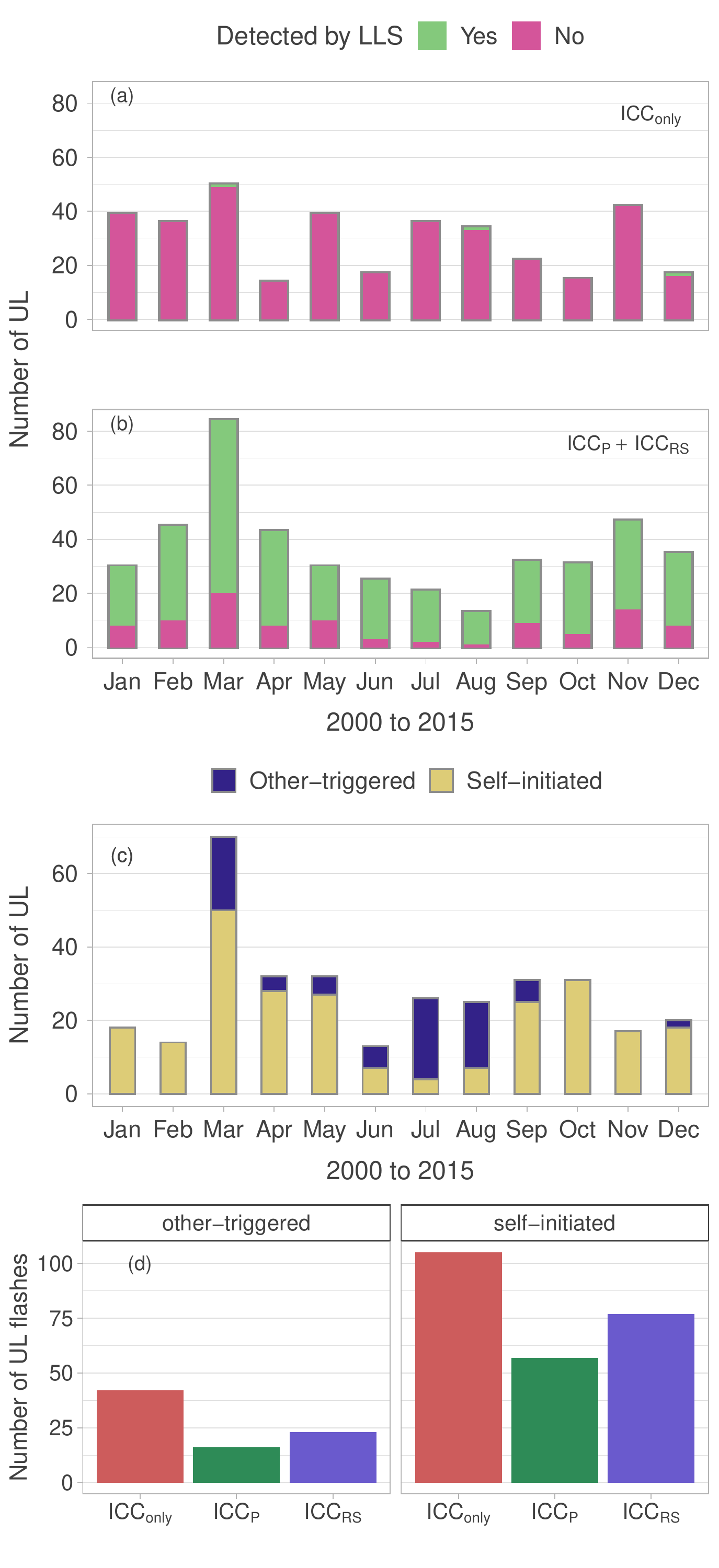}
\caption{Panel (a) and (b): Number of UL observed at the Gaisberg Tower (2000 to 2015: total of 792 UL) split into two categories. Upper panel (a): Initial continuous current only subtype (ICC\textsubscript{only}). Lower panel (b): Sum of UL with initial continuous current superimposed by pulses and by return-stroke sequences, respectively (ICC\textsubscript{P}\,+\,ICC\textsubscript{RS}). Colors distinguish detection (green) by the EUCLID lightning location system (LLS) from non-detection (red). Panel (c): Number of UL from Gaisberg Tower classified as self-initiated (yellow) and other-triggered (blue) following the classification scheme by \citeA{zhou2012}. Based on 329 observations from 2000 to 2015. Panel (d): Proportion of subtypes at Gaisberg Tower for other-triggered UL (left) and self-initiated UL (right).}
\label{fig:detect_trigg}
\end{center}
\end{figure}

One of the instrumented towers with a height of $100$~m is on top of the Gaisberg mountain in Austria. There, all lightning strikes are of the upward type \cite{diendorfer2015}. Figure~\ref{fig:detect_trigg} (a) shows that the ICC\textsubscript{only} subtype is basically undetectable by the regional LLS, whereas most of the other two subtypes (panel (b), with superimposed pulses ICC\textsubscript{P} or return strokes ICC\textsubscript{RS}) can be detected. While local, geographical and meteorological differences have been suspected influencing the occurrence of the ICC\textsubscript{only} UL type \cite{march2015, birkl2018, diendorfer2009}, to our knowledge no systematic search has been conducted. Finding widely available proxies from which its existence can be deduced would provide a better basis for proper risk assessment.

UL requires extremely high field intensities immediately above the ground or an object on it for a discharge to start \cite{rakov2003}. 
Structures with an effective height of $500$~m or more are assumed to deform the electric field sufficiently to experience only UL. For shorter objects, nearby lightning activity can deform the electric field sufficiently to trigger a so-called ``other-triggered'' upward leader. However, UL can also be initiated from the tall structure itself (``self-initiated'').

Location, e.g., being situated on an isolated hill, and meteorological conditions, are assumed to be favorable ingredients for self-initiated UL but details are not clear yet. Regional and seasonal differences are large. During six warm seasons at tower locations in the USA and Brazil only other-triggered UL occurred \cite{schumann2019}. In other locations and for other studies the ratio between self-initiated and other-triggered UL varies widely: $1$ to $1$ in Japan in six winter seasons \cite{wang2012}. In the USA, \citeA{warner2012} show that the ratio is $1$ to $4$; almost $80$~\% of self-initiated lightning and $15$~\% of other-triggered lightning occurred in the cold season. In Germany from measurements at the Peissenberg Tower, \citeA{manhardt2012} infer a ratio of $9$ to $1$; all self-initiated UL occurred in the cold season.  At the Gaisberg Tower, the ratio is $3$ to $1$ as shown in Fig.~\ref{fig:detect_trigg} (c). 

In addition to site-specific conditions \cite<e.g.,>[]{smorgonskiy2015}, differences in meteorological conditions have been proposed to explain variations in this ratio \cite{wang2012,jiang2014,zhou2014,smorgonskiy2015,yuan2017,mostajabi2018,pineda2019}. The results have been partly contradictory. For example, \citeA{zhou2014} and \citeA{smorgonskiy2015}, found no significant relationship between the ambient wind speed and self-initiated UL at Gaisberg, while \citeA{mostajabi2018} underline its importance both at Gaisberg and at S\"antis (Switzerland). Temperature has a significant impact at the Gaisberg Tower \cite<e.g.,>[]{zhou2014} and at the S\"antis Tower \cite<e.g.,>[]{smorgonskiy2015,pineda2019}, whereas wind speed is more crucial in studies conducted in Japan \cite{wang2012}, China \cite{yuan2017} or in the United States \cite{warner2014}.

Improving the understanding of the initiation mechanisms of UL and the occurrence of the LLS undetectable flash type from a general meteorological perspective might be a first step towards a proper assessment of the risk of UL to tall structures.

The major objective of this study is to assess whether a comprehensive set of larger-scale atmospheric variables may improve the understanding when UL is self-initiated and when UL is LLS undetectable.
Recent advances make the objective of this study achievable:  the availability of hourly and vertically highly resolved reanalyses of atmospheric conditions \cite<ERA5,>[]{hersbach2020} and powerful and flexible machine learning techniques with which to combine them with the lightning measurements.

This article is organized as follows: First, a brief overview of the data used is given including lightning observations and meteorological reanalysis data (section \ref{sec:data}).
Next a qualitative overview of influential atmospheric variables for self-initiated over other-triggered UL and ICC\textsubscript{only} over ICC\textsubscript{P}\,+\,ICC\textsubscript{RS} UL is given in Sect.~\ref{sec:trigger_quali} and \ref{sec:CC_quali}.
To quantify the influence of various atmospheric variables and the ability of these atmospheric variables to explain the occurrence of self-initiated UL and undetectable ICC\textsubscript{only} UL, the statistical approaches for the two stated issues are introduced in Sect.~\ref{sec:methods}. 
The quantitative results on the drivers for self-initiated over other-triggered UL and ICC\textsubscript{only} over ICC\textsubscript{P}\,+\,ICC\textsubscript{RS} UL are presented and discussed in Sect.~\ref{sec:results}. Finally, Sect.~\ref{sec:conclusion} summarizes the major findings.

\section{Data}\label{sec:data}

The study combines three different data sources. It uses UL data measured directly at the Gaisberg Tower in Salzburg (Austria), LLS data measured remotely by the European Cooperation for Lightning Detection \cite<EUCLID,>[]{schulz2016} and meteorological reanalysis data \cite<ERA5,>[]{hersbach2020}. In addition two variables are introduced to complement the climatological background reflecting atmospheric differences among seasons and among daytime. These are the day of the year and the hour of day.

\subsection{Lightning Observations}\label{sec:lightningdata}

Since 1998, UL has been directly measured at the Gaisberg Tower in Salzburg \cite<Austria,>[]{diendorfer2009}.
The $100$~m radio tower is situated on top of the Gaisberg mountain 1\,288 meters above mean sea level ($47 ^\circ 48 '$ N, $13 ^\circ 60'$ E). The study uses observations from 2000 to 2015. In total 819 UL flashes were recorded at the Gaisberg Tower during this period. 

The identification of self-initiated versus other-triggered UL flashes as well as the identification of the UL flash type relies on direct measurements at the Gaiserg Tower. Contrarily, the identification of nearby lightning relies on remote measurements by the LLS EUCLID.

Out of $819$ UL flashes, $329$ were assigned to be either self-initiated or other-triggered. $329$ UL events were investigated with respect to the ambient electrical field during initiation. If a transient field was present during the initiation at the Gaisberg Tower, the event was flagged as other-triggered lightning. If not, it was flagged as self-initiated UL. 
The assignment to one or the other initiation mechanism within these field measurement studies had therefore neither a temporal nor a spatial constraint. 

 $792$ UL observations could be unambiguously assigned to an UL type. $373$ UL flashes were classified as ICC\textsubscript{only} and $419$ UL flashes were classified as ICC\textsubscript{P}\,+\,ICC\textsubscript{RS} type UL. On the tower, a current-viewing shunt resistor allows to measure the overall current waveform of an UL flash (see \citeA{diendorfer2009} for details). While all UL flashes develop an ICC (initial continuous current), the overall waveform can be differentiated by three different characteristics leading to three distinct UL types. An ICC\textsubscript{RS} UL flash is characterized by a short phase after the ICC with no current followed by one or more return stroke like-sequences with more than $2$ kA. An ICC\textsubscript{P} UL flash is superimposed by one or more pulses with more then $2$ kA. ICC\textsubscript{only} type UL does not have pulses nor return strokes and has amplitudes lower than $2$ kA.

For the identification of nearby lightning discharges, EUCLID measures downward lightning with a detection efficiency of more than $90$~\% \cite{schulz2005}. The LLS measurements are used to determine whether UL was associated with nearby lightning activity. Other than for the procedure assigning an initiation mechanism, a temporal and spatial constraint is put on the UL event striking the Gaisberg Tower.

\subsection{Atmospheric Reanalysis}\label{sec:ERA5data}
ERA5 is ECMWF's fifth generation of global climate reanalysis from 1950 onward. ERA5 has a spatial resolution of $31$~km horizontally (available at a $0.25$~$^\circ$ $\times$ $0.25~^\circ$ latitude-longitude grid) and $137$ levels vertically at hourly resolution. This study considers the lowest $74$ levels extending to approximately $15$~km altitude, well into the stratosphere.

Convection results from various processes acting on different scales down to microscales which reanalysis data are not able to resolve. Nevertheless this study aims to find drivers for the UL type and initiation mechanism, however, from larger-scale meteorology. For this purpose the systematic search of atmospheric variables is not restricted to convection proxies but expanded to a comprehensive set of variables to involve various processes that might be influential. Thus in total 75 directly available and derived variables at the surface, on model levels and integrated vertically are spatially and temporally bilinearly interpolated to each Gaisberg Tower observation.
The atmospheric variables fall into five broad categories: cloud physics, temperature field, moisture field, surface exchange and wind field. Grouping the large set of variables into broad subgroups shall facilitate the interpretation. A complete list of the variable groups and individual variables can be found in the supporting information file.

\section{Explorative Analysis}\label{sec:qualitative}
Atmospheric conditions might play a crucial role in determining whether the tall structure of Gaisberg Tower itself initiates UL and whether ICC\textsubscript{only} UL occurs. Exploring the larger-scale meteorological conditions gives insights into the relationship between atmospheric variables and the initiation mechanism or the UL flash type. 

\subsection{Atmospheric Variables Influencing the Initiation Mechanism}\label{sec:trigger_quali}

The ability to distinguish conditions preferring self-initiated and other-triggered UL varies among different variables.

\begin{figure}
\begin{center}
\includegraphics[width=8.3cm]{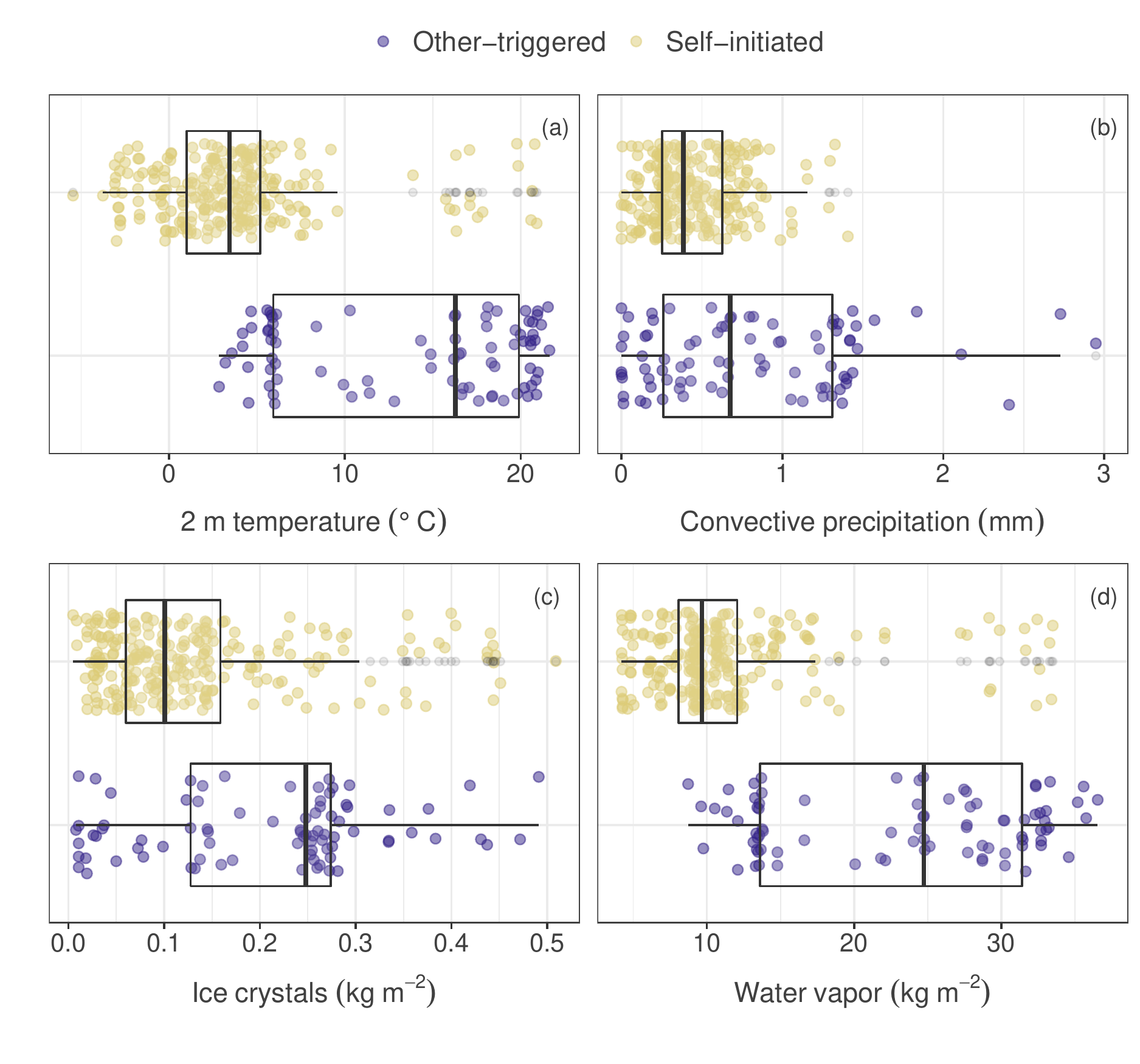}
\caption{Distribution of relevant atmospheric variables with respect to to self-initiated UL observations (yellow) and other-triggered UL observations (blue). Panels (a)--(d) illustrate the $2$~m temperature, convective precipitation, the total column ice crystals and the total column water vapor. Dots are spatially and temporally interpolated to Gaisberg observations ($329$) and jittered vertically for better visualization. Boxplots summarize the statistical distribution showing the median (vertical middle line), the interquartile range (IQR, $25$~\% to $75$~\%) and whiskers ($\pm$ $1.5~\cdot$ IQR).}
\label{fig:dist_points}
\end{center}
\end{figure}

Qualitatively, the $2$~m temperature and water vapor may most clearly distinguish self-initiated from other-triggered UL (Fig.~\ref{fig:dist_points}).
Self-initiated UL occurs at lower temperatures and lower amounts of water vapor. Further, it occurs at lower amounts of convective precipitation and ice crystals. 

Since temperature-related variables generally have a significant seasonal cycle, exploring the difference in the climatological distributions of some influential variables for one or the other initiation mechanism gives further insights.

\begin{figure}
\begin{center}
\includegraphics[width=8.3cm]{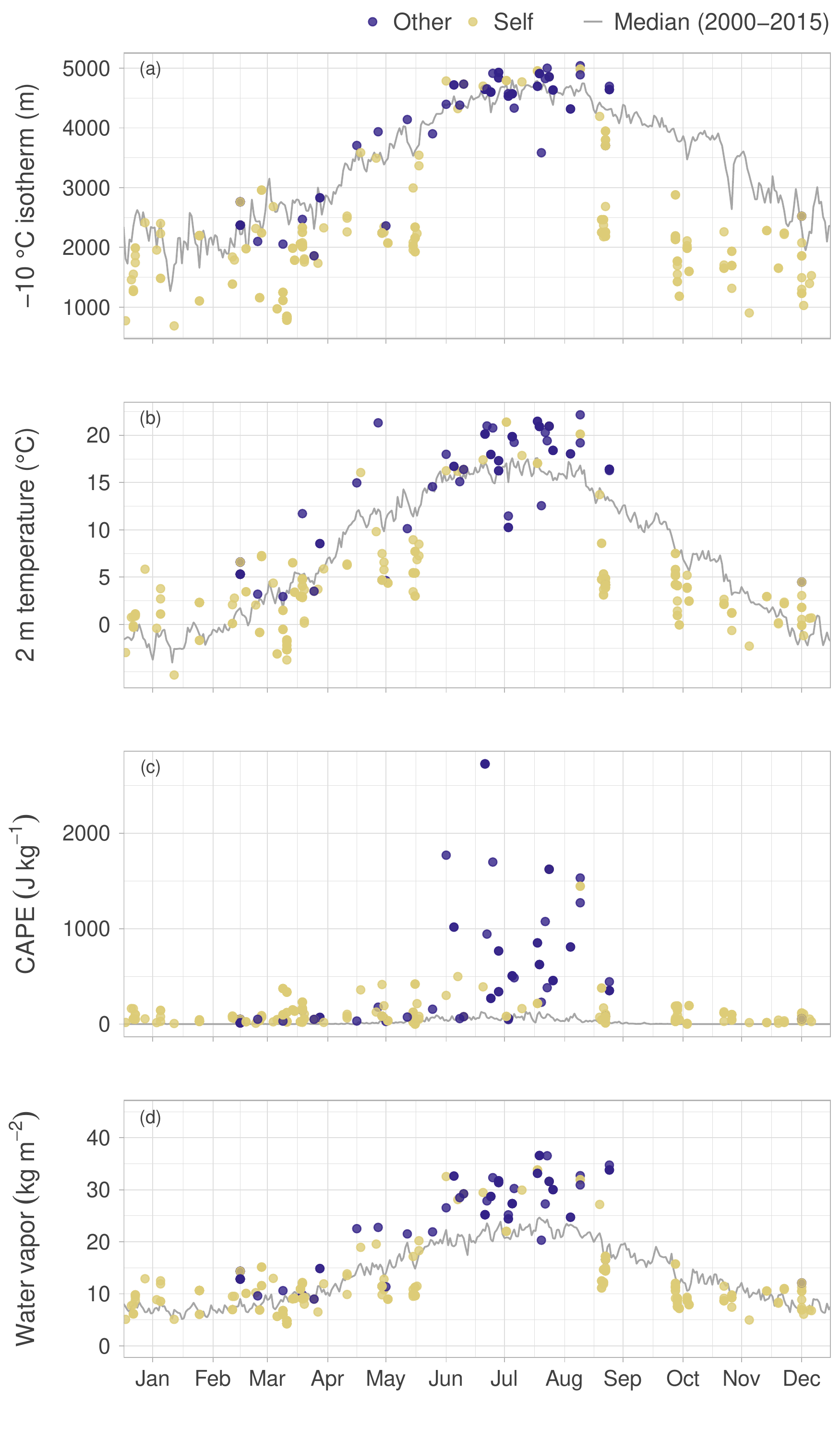}
\caption{Difference of self-initiated (yellow) and other-triggered (blue) conditions illustrated for four variables as deviation from the daily 2000 to 2015 Gaisberg medians (gray lines). From top to lowest panel (a)--(d):  $-10~^\circ$C  isotherm height above ground, $2$~m temperature, CAPE and total column water vapor. Dots indicate values spatially and temporally interpolated to UL events at the Gaisberg Tower (329 UL).}
\label{fig:matrixplot}
\end{center}
\end{figure}

Figure \ref{fig:matrixplot} illustrates that potentially relevant variables for self-initiated UL barely show an annual cycle, whereas other-triggered UL clearly are above the daily median from $2000$ to $2015$. To demonstrate this, three representatives from the temperature field (panel (a):  $-10~^\circ$C  isotherm height, (b): $2$~m temperature and (c): CAPE and (d): one representative from the moisture field (total column water vapor) are explored. Shown are the interpolated daily median values of the four different variables and deviations from these at UL initiation time. 
Self-initiated UL occurs preferentially associated with lower heights of the $-10~^\circ$C isotherm than the climatological median, especially during winter, spring and fall.
Similarly, self-initiated UL occurs at lower $2$~m temperature throughout the year but especially in the transition seasons. 
Moreover, other-triggered UL occurs associated with high CAPE and a high amount of water vapor throughout the year.
Hence results show that the specific influence of some variables is independent of the season. 

Results from the qualitative exploration are in line with findings by other studies. 
Most frequently the importance of temperature is explained by its influence on the height of the cloud charge layer. Lower temperatures lower the cloud base and bring the main electrification region closer to the tall structure \cite{heidler2014}. The reduced distance between the tall structure and the charged thundercloud then increases the electric field and eases the incidence of self-triggered lightning flashes \cite<e.g.,>{heidler2013,pineda2018,pineda2018b}.

\subsection{Atmospheric Variables Influencing the UL Flash Type}\label{sec:CC_quali}

\begin{figure}
\begin{center}
\includegraphics[width=8.3cm]{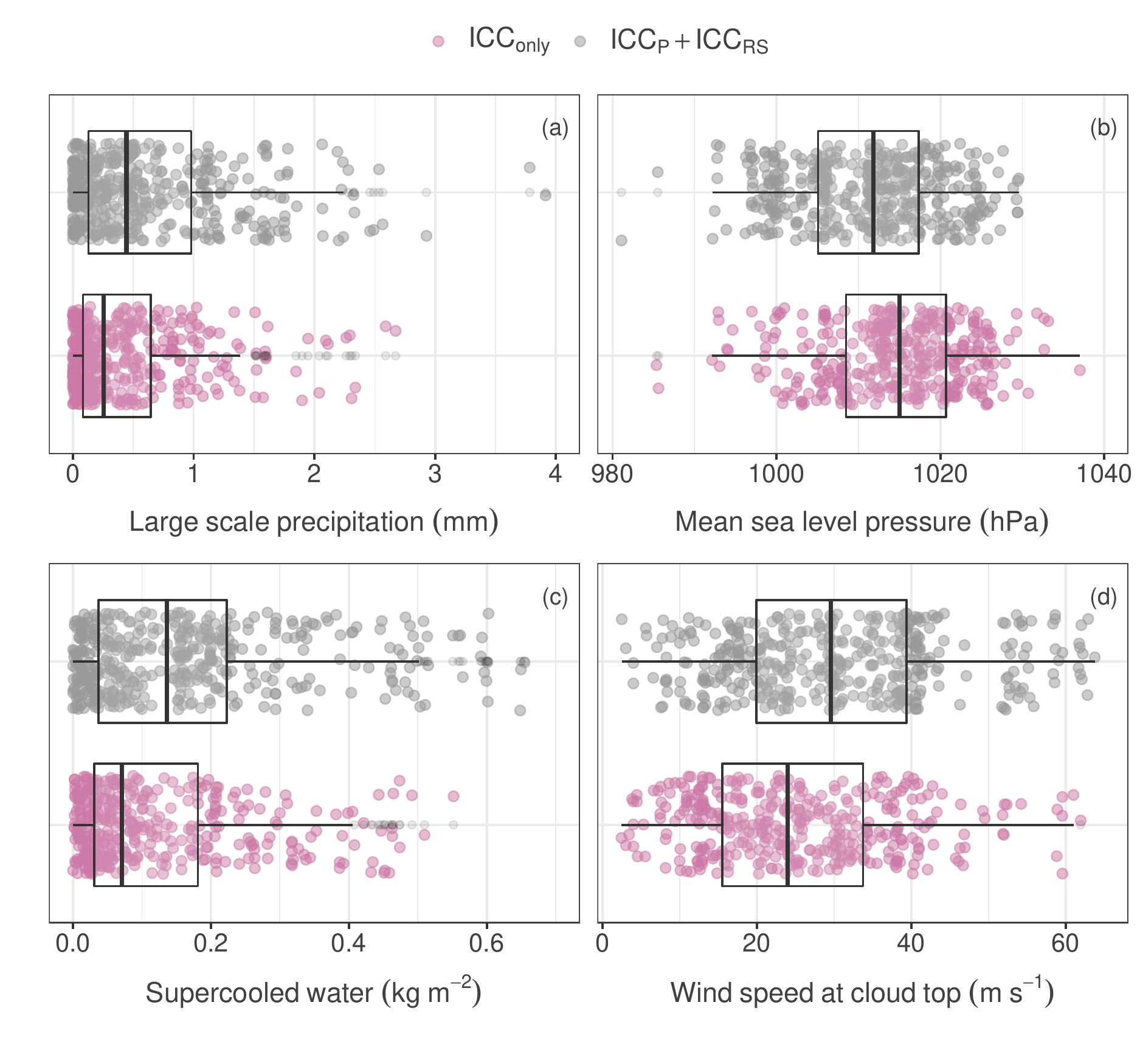}
\caption{Distribution of relevant atmospheric variables with respect to ICC\textsubscript{only} type UL (pink) and ICC\textsubscript{P}\,+\,ICC\textsubscript{RS} type UL (gray).  From top to lowest panels (a)--(d) illustrate large scale precipitation, mean sea level pressure, the total column supercooled liquid water and the wind speed at cloud top. Dots are spatially and temporally interpolated to Gaisberg observations ($792$) and jittered vertically. Boxplots summarize the statistical distribution.}
\label{fig:dist_points_type}
\end{center}
\end{figure}

In contrast to the initiation mechanism, an explorative analysis of atmospheric variables influencing the occurrence of one UL type over the other types does not yield clear results. 
The distributions of the four variables divided into the two flash type categories are very similar indicating a weaker ability for a clear classification (Fig.~\ref{fig:dist_points_type}). Still, some information from larger-scale meteorological variables may be inferred. 
The lower the amount of large scale precipitation, the proportion of supercooled liquid water, the wind speed at cloud top and the higher the mean sea level pressure, the more ICC\textsubscript{only} type UL is observed at the Gaisberg Tower according to the medians in the boxplots.

A qualitative analysis gives first insights and tendencies of variables having the ability to distinguish situations with self-initiated or other-triggered UL and ICC\textsubscript{only} type or ICC\textsubscript{P}\,+\,ICC\textsubscript{RS} type UL. However, it does not provide measurable results. 

The following sections introduce approaches to quantify (i) how well larger-scale meteorology is capable of explaining the initiation mechanism and UL flash type, (ii) which atmospheric variables drive the occurrence of a preferred initiation mechanism and UL flash type and (iii) the effect of influential variables on the initiation mechanism and UL flash type.

\section{Methodology of the Quantitative Analysis}\label{sec:methods}

The initiation mechanism and UL type occurrence might be highly nonlinearly related to the atmosphere represented by the $75$ atmospheric variables plus the two climatological background variables. To account for this relationship and to potentially compensate for the drawback of relatively coarse resolution of reanalysis data requires more sophisticated methods.

Regression and classification trees \cite{breiman1984} are popular models due to their flexibility and intuitive interpretation. However, they are also rather instable under small modifications of the learning data and often outperformed in terms of diagnostic performance. Using an ensemble of many trees on random subsamples of the full input data overcomes both of these problems, leading to random forests \cite{breiman2001} that are still extremely flexible and easy to set up while being more stable and typically with high diagnostic ability.

Ensembles of classification trees, i.e., random forests, shall capture the potential nonlinearities and interactions in the relationship between the two response variables and the explanatory variables. More specifically, conditional inference random forests are used (section \ref{sec:decisiontrees_rf}) combined with permutation-based variable importance measures to assess the relative influence of the different atmospheric drivers (section \ref{sec:permvarimp}). Random forests are a powerful machine learning technique used in various scientific fields \cite{strobl2009}.

\subsection{Statistical Modeling Through Random Forests}\label{sec:decisiontrees_rf}

Finding the drivers for the initiation mechanism and the UL type are binary classification problems. For the initiation mechanisms, it is self-initiated versus other-triggered UL; for the UL type, the response variable is ICC\textsubscript{only} versus ICC\textsubscript{P}\,+\,ICC\textsubscript{RS} type UL. Seventy-seven variables (cf. section \ref{sec:ERA5data}) are considered as explanatory variables.

Conditional inference random forests \cite{hothorn2004} are employed consisting of individual trees \cite{hothorn2006}. Each tree consists of nodes which are represented by so-called split variables chosen based on permutation tests \cite<also known as conditional inference,>[]{strasser1999}. Growing these individual trees comprises three steps. Finding an appropriate split variable from the set of atmospheric variables, finding an appropriate threshold where to split in the respective split variable and finding a point where to stop the tree growing. A simple example of a classification tree can be found in the supplementary information file.

Each of these trees is based on a random subsample of the input data and is constructed in the following way.
First the atmospheric variable with the strongest association is selected as split variable. Next a reasonable split point in this split variable to separate the different response classes as well as possible is found using a permutation test statistic. Computing this test statistic over all possible subsets and setting the split point where it is most reasonable according to the test statistic, allows to judge which threshold adds the most to the performance. The same driver and split point selection is then repeated recursively for all of the random subsamples of the input data.
Splitting continues until a certain stopping criterion (e.g., significance or subsample size) is met. The random forest then averages the diagnosed probabilities from the ensemble of trees, which stabilizes and enhances the diagnostic performance.
More details regarding the algorithm and corresponding implementation are provided in \citeA{hothorn2006} and \citeA{hothorn2015}.

\subsection{Assessing the Variable Importance}\label{sec:permvarimp}

Random forests greatly stabilize the inferred regression relationship and improve the diagnostic performance. 
One of the largest benefits using random forests is the ability to assess the so-called variable importance.

Here this approach is followed and employed by the permutation variable importance \cite<see>[]{strobl2008}. In short, the idea is to break up the relationship between the response variable and one driver variable by permuting, i.e., randomly mixing its values of, the latter and then assessing how much the diagnostic performance deteriorates.

To assess this diagnostic performance the variable importance is computed on data not used for modeling. The random forests are learned on two thirds of the input data and the permutation variable importance is assessed on the remaining one-third of the input data (i.e., test data).

Diagnostic performance on the test data is described by a specific metric (score). In this study it is the area under the receiver operating characteristic curve \cite<AUC,>[]{wilks}. AUC is a scalar value representing the ability to distinguish and hence correctly classify the response. The closer this value is to $1$ the higher the performance. For the variable importance a single driver variable is randomly permuted and the median AUC decrease, i.e., the decrease in performance, is computed across all test data sets. This decrease of diagnostic performance when permuting the indicated variable is computed by  the normalized difference of the original sample score and the permuted sample score. The driver variable leading to the strongest decrease in the diagnostic performance metric, i.e., AUC, is most important for the classification.

\section{Model Results and Discussion}\label{sec:results}
Random forest models classify self-initiated and other-triggered flashes and ICC\textsubscript{only} and ICC\textsubscript{P}\,+\,ICC\textsubscript{RS} from the $75$ larger-scale meteorological variables plus the two additional background variables (time of day and day of year). 

\subsection{Atmospheric Influence and Drivers for the Initiation Mechanism and the UL flash type}

The random forest models can reliably differentiate between self-initiated and other-triggered UL from the $77$ variables. The area under the curve (AUC) is at $0.93$ very close to the optimum value of $1$. Additionally the separation is both reliable and sharp \cite{gneiting2007}. A more detailed description can be found in the supplementary information file.

\begin{figure}
\begin{center}
\includegraphics[width=11cm]{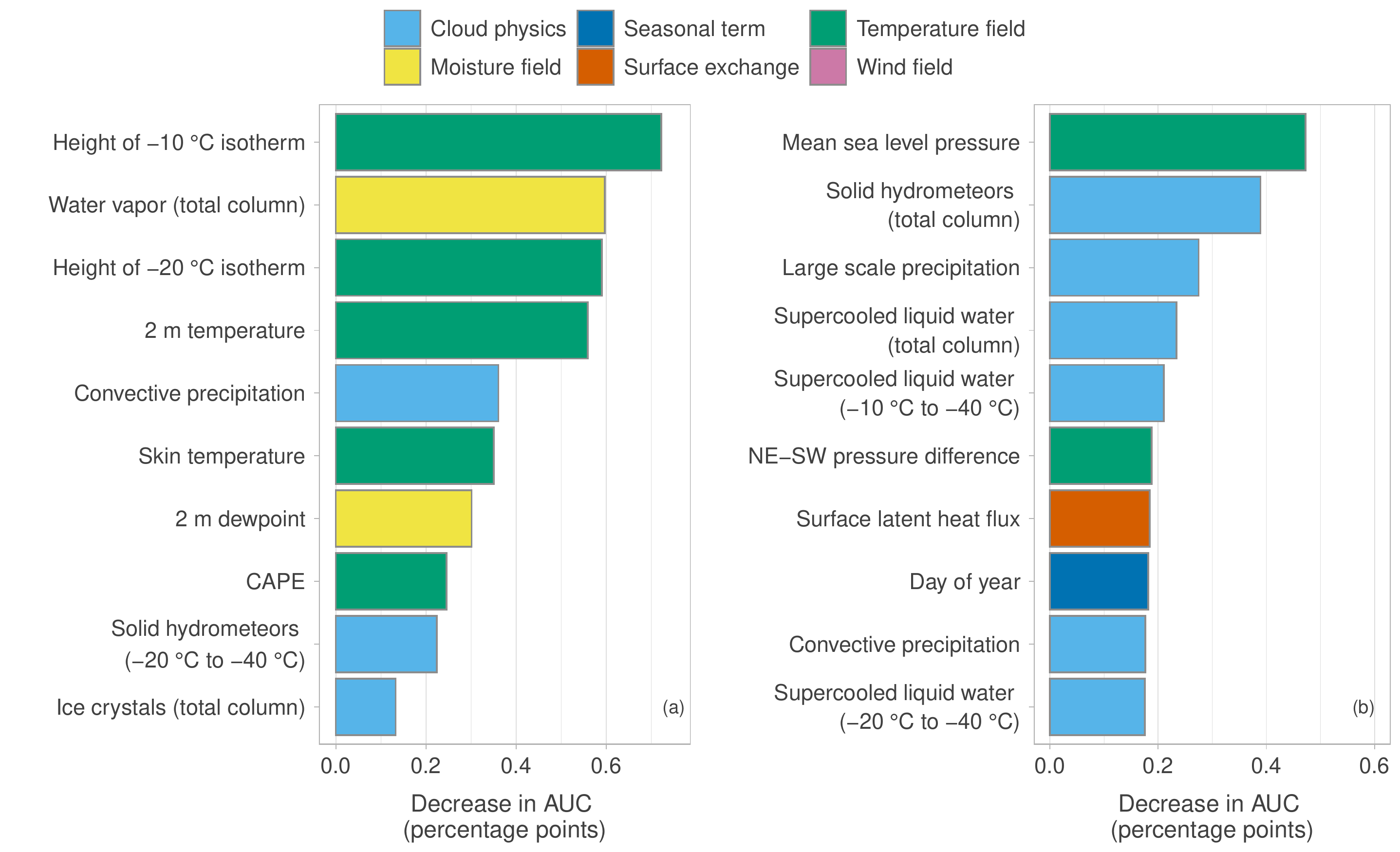}
\caption{Ranking of the most important variables influencing whether (a) UL is self-initiated and (b) is of ICC\textsubscript{only} type according to the permutation variable importance procedure described in section \ref{sec:permvarimp}. Colors indicate to which meteorological group the variables belong.}
\label{fig:varimp_ind}
\end{center}
\end{figure}

The random forest models together with the variable importance can clearly identify which variables are most important for the differentiation of the initiation mechanisms. 
The importance of the driving variables for the initiation mechanism is ranked in Figure~\ref{fig:varimp_ind} (a). Five of the most important variables listed are part of the temperature field group. Particularly the height of the $-10~^\circ$C isotherm has a large influence on the self initiation of UL. Further the  $-20~^\circ$C  isotherm height, the $2$~m temperature, CAPE and the skin temperature have an impact. Two additional variables, strongly related to the temperature field are the total column water vapor and the $2$~m dewpoint temperature from the moisture field group. 
The most influential variables from cloud physics are the convective precipitation, the proportion of solid hydrometeors between  $-20~^\circ$C  and  $-40~^\circ$C  and ice crystals (total column). 

The temperature-related variables  $-10~^\circ$C  and  $-20~^\circ$C  isotherm height, $2$~m temperature, skin temperature, $2$~m dewpoint temperature and water vapor in Fig.~\ref{fig:varimp_ind} all share relatively high mutual information values. 
Following the model fitting and the permutation variable importance process described in Sect. \ref{sec:methods}, different variables may be selected as potential splitting variables at different stages. If in one stage the most important variable, i.e., the $-10~^\circ$C  isotherm height is not selected, it might be replaced by a variable serving as proxy for it.
This highlights that the random forest models require any information on the distance between the tall structure and the main electrification region which is closely related to any temperature information.

For the UL flash type classification, the area under the curve is only $0.66$ indicating that larger-scale meterological information used as input for the random forests does not carry enough information on the distinction between ICC\textsubscript{only} UL and the the other two subtypes of UL.

Investigating the permutation variable importance yields results in line with the explorative analysis in Sect.~\ref{sec:CC_quali}.
Cloud physics variables dominate the list of the ten most influential variables when they are individually randomly permuted. Six out of the ten are cloud physics variables (see panel (b) in Fig.~\ref{fig:varimp_ind}). 
Permuting individual variables shows that the mean sea level pressure has the largest influence on the probability of ICC\textsubscript{only} UL over ICC\textsubscript{P}\,+\,ICC\textsubscript{RS} UL. Further, the N-E and S-W ($1.4~^\circ$) pressure difference of the Gaisberg Tower, the surface latent heat flux and the day of the year (i.e., the season) have a comparable impact in magnitude on the probability of ICC\textsubscript{only} UL.

\begin{figure}
\begin{center}
\includegraphics[width=8.3cm]{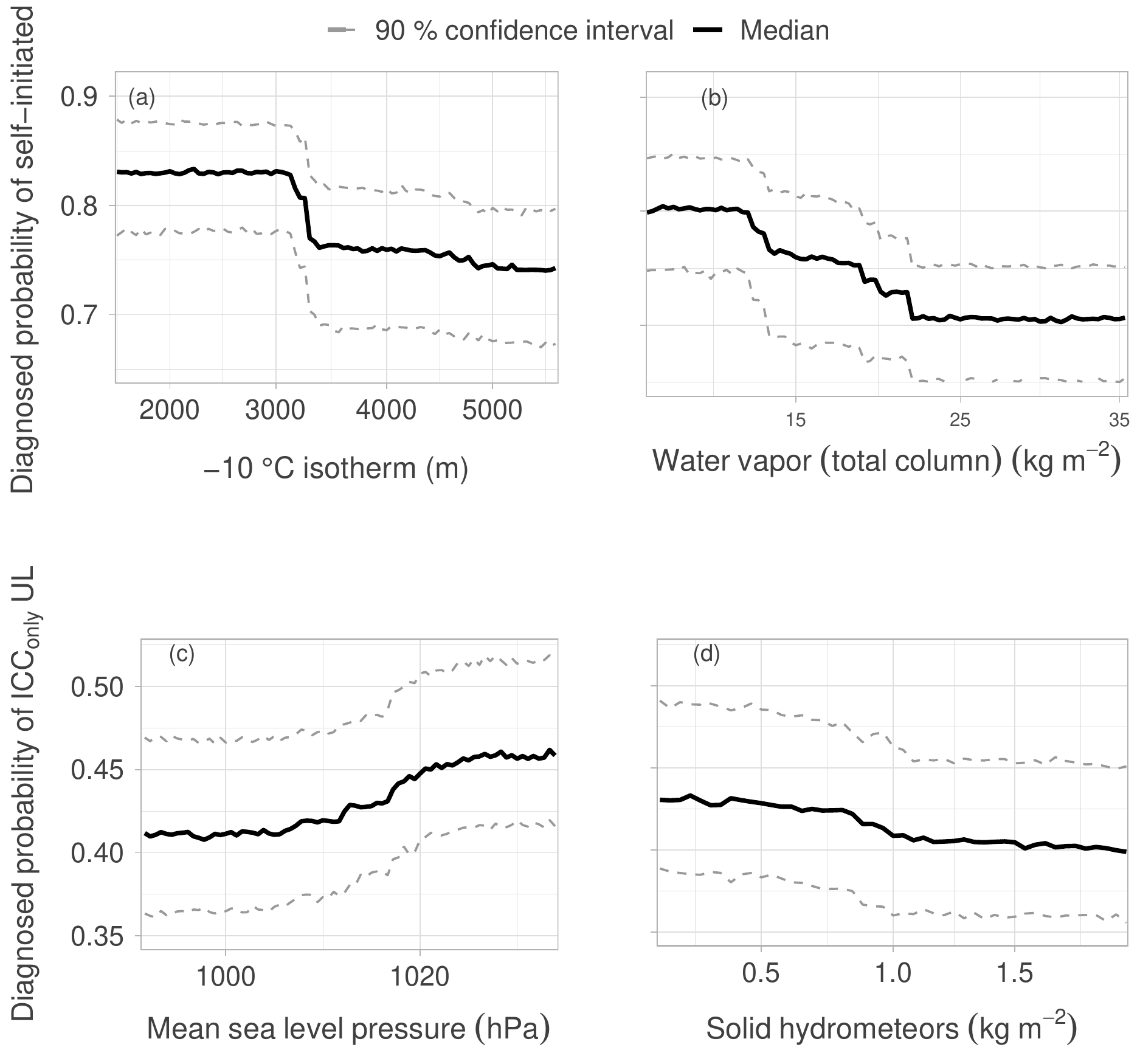}
\caption{Left (a): Effect of the  $-10~^\circ$C  isotherm height above ground on probability of self-initiated UL (value $1$). Right (b): Effect of total column water vapor on probability of self-initiated UL. 
Lower left (c): Effect of mean sea level pressure on probability of ICC\textsubscript{only} UL (value $1$). Lower right (d): Effect of solid hydrometeors (total column). Effect as solid lines are diagnosed median. Dashed lines are lower and upper bound of the $90$~\% confidence interval indicating the uncertainty of the random forests.}
\label{fig:effects_CC_ind}

\end{center}
\end{figure}

The second important benefit of the conditional inference random forest models is that the effect of a single variable on the probability of self-initiated UL or ICC\textsubscript{only} UL can be assessed.
 The influence of a single variable in the random forests can be explained by varying it while other explanatory variables are held constant at their mean value.
Varying the $-10~^\circ$C isotherm height, Fig.~\ref{fig:effects_CC_ind} (a) illustrates, that a higher $-10~^\circ$C isotherm decreases the probability of self-initiated UL. The range between the dashed lines illustrates the confidence interval, i.e., the uncertainty of the effect. The solid line shows the median diagnosed effects from the ensembles based on $100$ learning samples including two thirds of the original number of observations. 
 Note a drop in the probability of self-initiated UL associated with a $-10~^\circ$C isotherm between around $3~000$~m and $3~500$~m above ground. This drop makes about $7$ percentage points in the probability of self-initiated UL according to the median based on the random forests. 
Varying the total column cloud water vapor shows that a higher amount decreases the probability of self-initiated over other-triggered UL (panel (b) in Fig.~\ref{fig:effects_CC_ind}). 

For the flash type, varying the mean sea level pressure shows that that a higher mean sea level pressure increases the probability of ICC\textsubscript{only} type UL (panel (c) of Fig.~\ref{fig:effects_CC_ind}). According to the median this probability increases from about $0.40$ to $0.46$ when the mean sea level pressure changes from $1~000$~hPa to $1~030$~hPa. 
Solid hydrometeors in panel (d) have the opposite effect on the probability of ICC\textsubscript{only} UL. A lower proportion of solid hydrometeors increases the probability of ICC\textsubscript{only} over the other subtypes of UL.
 The same is true for other variables such as the total column supercooled liquid water, the total column snow water or total column ice water (not shown here). A possible interpretation would be that in ICC\textsubscript{only} UL flash situations, convection may be less pronounced causing that less hydrometeors are built or hydrometeors are more short-lived. 
 
 The advantage using random forests is that the relatively low improvement of distinguishing the UL flash type of a single variable may be crucially improved by combining many different variables. However, for the flash type classification, larger-scale meteorological information can only provide limited information.

%
%
%

 Geographical and local topographical effects might have a significant influence on various UL parameters as emphasized by \citeA{march2015} and \citeA{birkl2018} which might also influence the ability of the meteorlogical setting to distinguish the UL types.

Beyond the larger-scale meteorological setting, processes on the convective scale certainly play a substantial role. However, these convective processes are not explicitly resolved in reanalysis data. Hence, information on the result of convective processes, e.g., indicated by the current natural lightning activity, might be worth to consider. 
Panel (d) in Fig.~\ref{fig:detect_trigg} shows that from direct field measurements the relative proportion of UL subtypes for self-initiated and other-triggered UL is very similar. This suggests that there is no explanatory power of the initiation mechanism for the different UL flash types.
In this context it might be interesting whether nearby discharges occur at all during UL initiation at Gaisberg and if so whether the closest activity is near or far away from the Gaisberg Tower.

\subsection{The Influence of Nearby Lightning Activity on the UL Flash Type}\label{sec:elphys}

An exploratory analysis of the relationship of nearby lightning activity and the UL flash type yields a clear result. The spatially and/or temporally closest discharge to an UL flash at the Gaisberg Tower within a window of $60$~s before and after an UL flash and a radial distance of $100$~km extracted from EUCLID LLS is used.

Only $36$~\% of ICC\textsubscript{only} type UL are accompanied by nearby lightning activity, but $89$~\% of the other two subtypes are accompanied by nearby lightning activity  (Tab.~\ref{table:tab}). This suggests the hypothesis that ICC\textsubscript{only} flashes occur preferably when conditions are not intense enough to produce strong natural downward lightning activity. This suggests that ICC\textsubscript{P}\,+\,ICC\textsubscript{RS} UL occur when conditions also favor natural downward lightning, whereas ICC\textsubscript{only} flashes occur when conditions are not favorable for downward lightning. 

\begin{table}[]
\centering
\caption{Relative proportion of UL events with and without discharge activity within $100$~km distance and $60$~s before and after initiation separated into the UL flash type categories.}
\label{table:tab}
\begin{tabular}{lr} \\
& \textbf{Nearby Lightning Activity} \\
\end{tabular}

\begin{tabular}{lll}
 &  Yes   & No \\ \hline \\
LLS undetectable  & $36 \%$              & $64 \%$          \\
LLS detectable      &$89 \%$   & $11 \%$     \\ \hline   
\end{tabular}
\end{table}

\begin{figure}
\begin{center}
\includegraphics[width=8.3cm]{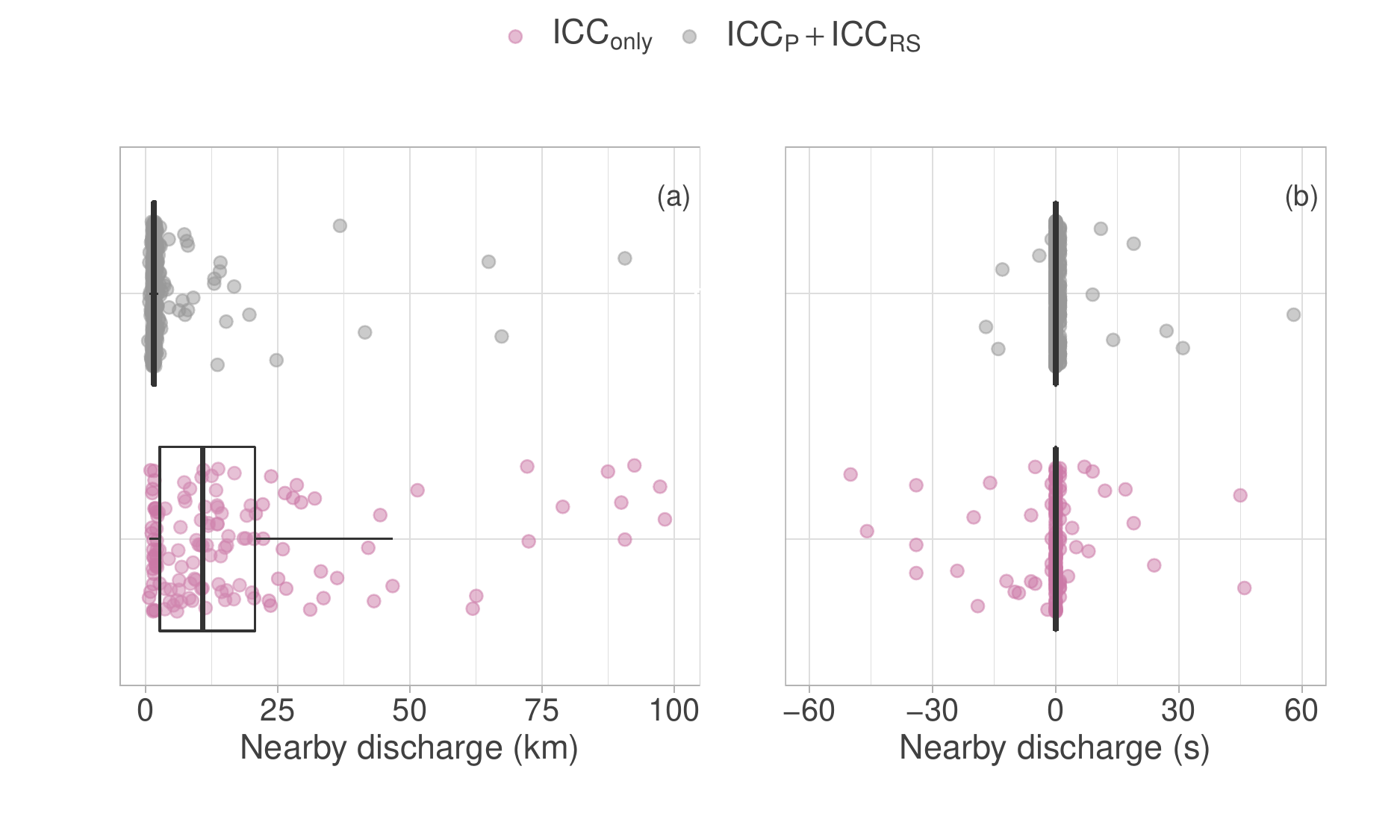}
\caption{Left (a): Distributions of distances of closest nearby lightning within $60$s before and after UL initiation at the Gaisberg Tower with respect to ICC\textsubscript{only} type UL (pink) and ICC\textsubscript{P}\,+\,ICC\textsubscript{RS} type UL (gray) observations. Right (b): Distributions of temporal distance of closest neraby lightning within $100$~km distance. Dots are scattered vertically for better visualization. Boxplots summarize the statistical distribution showing the median (vertical middle line), the interquartile range (IQR, $25$~\% to $75$~\%) and whiskers ($\pm$ $1.5~\cdot$ IQR).}
\label{fig:distance_points}
\end{center}
\end{figure}

For a better understanding of the interaction of nearby lightning discharges and the initiation mechanism and type of UL spatial and temporal distance of nearby lightning activity are explored when UL occurs at the Gaisberg Tower. 
Given that nearby lightning is present, Fig.~\ref{fig:distance_points} illustrates the distributions of the spatial (left) and temporal (right) distance of nearby discharges to the Gaisberg Tower with respect to the different UL flash type categories.

When ICC\textsubscript{P}\,+\,ICC\textsubscript{RS} type UL flashes occur the closest nearby lightning activity is rarely more than a few hundred meters away from the Gaisberg Tower. This suggests that these types of UL favor locally concentrated and potentially stronger convection in which both downward and upward lightning occur. The meteorological setting is hence both favorable for natural downward lightning and UL of the pulse and return stroke type initiated from the Gaisberg Tower.

Given that nearby discharges are around when ICC\textsubscript{only} type flashes occur, the closest activity is most often more than a few kilometers away. 
 The meteorological setting is hence neither favorable for natural downward lightning in the direct vicinity of the Gaisberg Tower nor for ICC\textsubscript{P}\,+\,ICC\textsubscript{RS} type UL flashes, however, favorable for the ICC\textsubscript{only} UL flash type.

In the following, random forest models quantify how important the nearby discharges are for the diagnosis which UL flash type most likely occurs.

Three variables are added to the set of explanatory variables: The spatial and the temporal distance of nearby discharges and the product of both. To account for the absence of nearby lightning activity and also the vicinity to the Gaisberg Tower, given that nearby lightning activity is present requires transforming the additional explanatory variables. Exponential kernel functions with a normalization of the radial distance by $100$~km, and temporal distance by $60$~s before and after an UL flash are applied to them. Values approach $1$ the closer nearby discharges to the UL in space and time and approach $0$ the further away from the Gaisberg Tower in space and time. A value of $0$ indicates that there is no nearby lightning activity within $100$~km and/or $60$~s before or after the UL flash.

Results from the random forest models show that the improvement from including these explanatory variables is large when assessing the diagnostic performance. Including them yields an AUC of $0.9$ showing a high diagnostic ability. In other words, including the nearby discharge activity allows to reliably separate the two categories of UL flash type.

%
%

\begin{figure}
\begin{center}
\includegraphics[width=8.3cm]{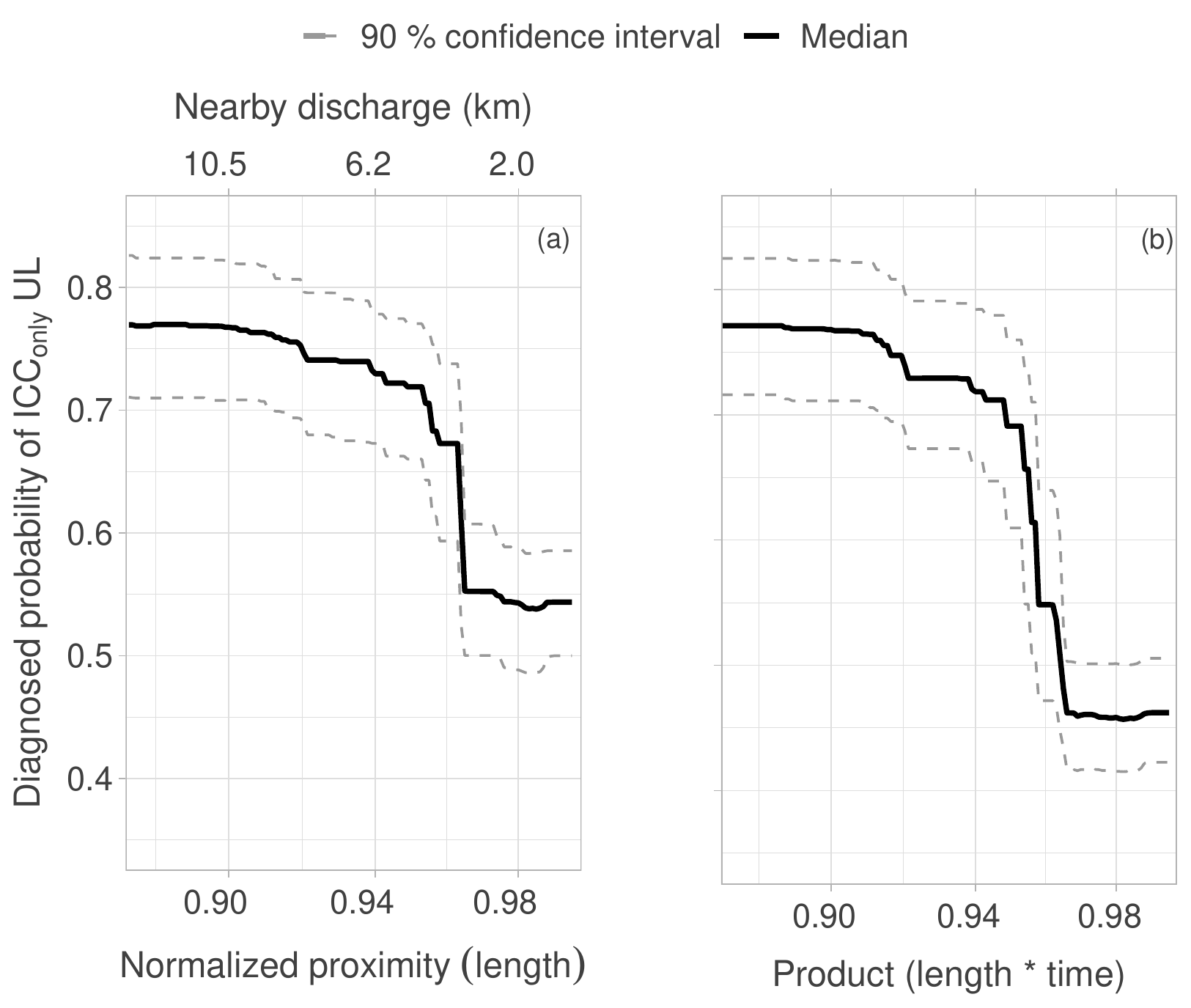}
\caption{Left (a): Effect of discharge distance in space on probability of ICC\textsubscript{only} UL (value $1$). Center (b): Effect of nearby discharge distance in time. Right (c): Effect of product of discharge distance in space and time. Effect as solid lines are diagnosed median. Dashed lines are lower and upper bound of the $90$~\% confidence interval indicating the uncertainty of the random forests.}
\label{fig:effects_elphys}
\end{center}
\end{figure}

The farther away (in space and time) the nearby discharge is, the higher the probability that the upward lightning flash is of type ICC\textsubscript{only} (Fig.~\ref{fig:effects_elphys}). The probability does not change gradually but jumps step-like by about $10$ percentage points at a discharge distance of about $4$~km (a). The product of spatial and temporal distance yields an even bigger step of about $20$ percentage points (b).
Formulated differently, the results in Fig.~\ref{fig:effects_elphys} show that lightning discharges close to location and time of upward lightning initiation favor the inception of ICC\textsubscript{P}\,+\,ICC\textsubscript{RS} flashes, whereas the absence of nearby lightning significantly increases the probability that an upward flash is of type ICC\textsubscript{only}. Nevertheless, however, UL depends on the presence of clouds and is thus not a fair-weather phenomenon.
\section{Summary and Conclusions}\label{sec:conclusion}

This study investigates qualitatively and quantitatively the larger-scale atmospheric impact on the initiation mechanism and upward lightning (UL) flash type at the Gaisberg Tower. 
It applies random forests, a powerful and flexible machine learning technique, to a climatological dataset of larger-scale atmospheric variables and lightning measurements to diagnose (i) whether UL initiated from a tall tower is triggered by nearby lightning (``other-triggered'') or by the tower itself (``self-initiated''), and (ii) of what particular subtype it is.

Measurements of UL at the Gaisberg Tower in Austria between 2000 and 2015 are combined with $75$ atmospheric variables derived from the ERA5 reanalysis, time of day, day of year, and the occurrence of nearby lightning discharges detected by the EUCLID lightning location system. 
Whether UL is self-initiated can be reliably explained by larger-scale atmospheric variables. The most important variable is the height of the $-10~^\circ$C isotherm. As the distance to the tall structure decreases, the probability of UL being self-initiated increases. Further important variables are the $-20~^\circ$C isotherm height, the $2$~m temperature and CAPE. An important contribution comes further from variables of the cloud physics group such as water vapor, convective precipitation and the proportion of solid hydrometeors between $-20~^\circ$C and $-40~^\circ$C and ice crystals. The lower the amount of water (gaseous, liquid or solid), the higher the probability of self-initiated UL.

Whether UL is of the ICC\textsubscript{only} type has important consequences because that type cannot be spotted by lightning location systems but only by specially instrumented towers so that no regionally detailed information of their occurrence exists. An ICC\textsubscript{only} type UL is characterized by an initial continuous current only, whereas the other types have superimposed pulses or return stroke(s).

For the occurrence of ICC\textsubscript{only} UL, larger-scale atmospheric conditions are far less influential than for the initiation mechanism. The random forest results clearly indicate that the meteorological setting derived from larger-scale atmospheric variables constitutes only one part and cannot fully explain the occurrence of ICC\textsubscript{only} UL. 

However, nearby (in space and time) lightning activity shows a strong relationship to the type of UL at the Gaisberg Tower. ICC\textsubscript{only} type UL occurs most often when there is no nearby lightning activity at all within $100$~km $60$~s before and after the UL flash initiated from the tower. Further, given that there is nearby lightning activity ICC\textsubscript{P}\,+\,ICC\textsubscript{RS} type UL favors conditions in which the activity is concentrated around the tower within only a few hundred meters distance. The farther away the nearby lightning activity the more favorable for ICC\textsubscript{only} type UL on the contrary.
This suggests that conditions (meteorological, topographical, electrical field) around the Gaisberg Tower that favor natural downward lightning also favor ICC\textsubscript{P}\,+\,ICC\textsubscript{RS} type UL. On the other hand conditions favoring ICC\textsubscript{only} type UL do not favor natural downward lightning.


\acknowledgments
We acknowledge the funding of this work by the Austrian Research Promotion
Agency (FFG), project no.~872656 and Austrian Science Fund (FWF) grant
no.~P\,31836. We thank Siemens BLIDS for providing EUCLID data.
The computational results presented here have been achieved in part using the LEO HPC infrastructure of the University of Innsbruck.

\subsection*{Author contribution}
Isabell Stucke did the investigation, wrote software,
visualized the results and wrote the paper. Deborah Morgernstern, Thorsten Simon and
Isabell Stucke performed data curation, built the data set, and derived
variables based on ERA5 data. Thorsten Simon contributed with coding concepts.
Georg J. Mayr provided support on the meteorological analysis, data organization and funding acquisition. Achim Zeileis supervised the formal analysis and
interpretation of the statistical methods. Achim Zeileis, Georg J. Mayr, and Thorsten Simon are the project administrators and supervisors. Gerhard Diendorfer, Wolfgang Schulz and Hannes Pichler made all lightning data available, this study relies on. Further they contributed to the interpretation of results from prior knowledge in lightning research. All authors contributed to the conceptualization of this paper, discussed on the
methodology, evaluated the results, and commented on the paper.

\section*{Open Research}
\subsection*{Data availability}
ERA5 data are freely available at the Copernicus Climate Change Service (C3S) Climate Data Store \cite{hersbach2020}. The results contain modified Copernicus Climate Change Service information (2020). Neither the European Commission nor ECMWF is responsible any use that may be made of the Copernicus information or data it contains.
EUCLID data and direct observations from the Gaisberg Tower are available only on request. For more details contact Wolfgang Schulz or Siemens BLIDS.

\subsection*{Software}\label{sec:app_software}    
All calculations as well as setting up the final data sets, modeling and the diagnosis were performed using R \cite{R}, using packages netCDF4 \cite{pkgncdf4},  partykit  \cite{hothorn2015}, ggplot2 package \cite{pkgggplot2}. Retrieving the raw data and deriving further variables from ERA5 required using Python3 \cite{python3} and cdo \cite{cdo}.

\subsection*{Competing interests}
The authors declare that they have no conflict of interest.

%
\bibliography{references} 

\end{document}